\documentstyle[11pt,epsfig]{article}
\newcounter{subequation}[equation]

\makeatletter 
\def\thesubequation{\theequation\@alph\c@subequation}
\def\@subeqnnum{{\rm (\thesubequation)}}
\def\slabel#1{\@bsphack\if@filesw {\let\thepage\relax
   \xdef\@gtempa{\write\@auxout{\string
      \newlabel{#1}{{\thesubequation}{\thepage}}}}}\@gtempa
   \if@nobreak \ifvmode\nobreak\fi\fi\fi\@esphack}
\def\subeqnarray{\stepcounter{equation}
\let\@currentlabel=\theequation\global\c@subequation\@ne
\global\@eqnswtrue
\global\@eqcnt\z@\tabskip\@centering\let\\=\@subeqncr
$$\halign to \displaywidth\bgroup\@eqnsel\hskip\@centering
  $\displaystyle\tabskip\z@{##}$&\global\@eqcnt\@ne
  \hskip 2\arraycolsep \hfil${##}$\hfil
  &\global\@eqcnt\tw@ \hskip 2\arraycolsep
  $\displaystyle\tabskip\z@{##}$\hfil
   \tabskip\@centering&\llap{##}\tabskip\z@\cr}
\def\endsubeqnarray{\@@subeqncr\egroup
                     $$\global\@ignoretrue}
\def\@subeqncr{{\ifnum0=`}\fi\@ifstar{\global\@eqpen\@M
    \@ysubeqncr}{\global\@eqpen\interdisplaylinepenalty \@ysubeqncr}}
\def\@ysubeqncr{\@ifnextchar [{\@xsubeqncr}{\@xsubeqncr[\z@]}}
\def\@xsubeqncr[#1]{\ifnum0=`{\fi}\@@subeqncr
   \noalign{\penalty\@eqpen\vskip\jot\vskip #1\relax}}
\def\@@subeqncr{\let\@tempa\relax
    \ifcase\@eqcnt \def\@tempa{& & &}\or \def\@tempa{& &}
      \else \def\@tempa{&}\fi
     \@tempa \if@eqnsw\@subeqnnum\refstepcounter{subequation}\fi
     \global\@eqnswtrue\global\@eqcnt\z@\cr}
\let\@ssubeqncr=\@subeqncr
\@namedef{subeqnarray*}{\def\@subeqncr{\nonumber\@ssubeqncr}\subeqnarray}
\@namedef{endsubeqnarray*}{\global\advance\c@equation\m@ne%
                           \nonumber\endsubeqnarray}

\makeatletter
\@addtoreset{equation}{section}
\makeatother
\renewcommand{\theequation}{\thesection.\arabic{equation}}
\def\dalemb#1#2{{\vbox{\hrule height .#2pt
        \hbox{\vrule width.#2pt height#1pt \kern#1pt
                \vrule width.#2pt}
        \hrule height.#2pt}}}
\def\square{\mathord{\dalemb{6.8}{7}\hbox{\hskip1pt}}}

    \let\e=\epsilon
  \let\q=\theta  
  \let\n=\nu

 \def\bd{\begin{document}} \def\ed{\end{document}}
\def\ds{\documentstyle} \let\fr=\frac \let\bl=\bigl \let\br=\bigr
\let\Br=\Bigr \let\Bl=\Bigl 
\let\bm=\bibitem
\let\na=\nabla
\let\pa=\partial \let\ov=\overline
\def\ie{{\it i.e.\ }} 
\newcommand{\be}{\begin{equation}} 
\newcommand{\ee}{\end{equation}} 
\def\ba{\begin{array}}
\def\ea{\end{array}}
\def\ft#1#2{{\textstyle{{\scriptstyle #1}\over {\scriptstyle #2}}}}
\def\fft#1#2{{#1 \over #2}}
\def\del{\partial}
\def\sst#1{{\scriptscriptstyle #1}}
\def\oneone{\rlap 1\mkern4mu{\rm l}}
\def\e7{E_{7(+7)}}
\def\td{\tilde}
\def\wtd{\widetilde}
\def\im{{\rm i}}
\def\bog{Bogomol'nyi\ }
\def\q{{\tilde q}}
\def\hast{{\hat\ast}}
\def\0{{\sst{(0)}}}
\def\1{{\sst{(1)}}}
\def\2{{\sst{(2)}}}
\def\3{{\sst{(3)}}}
\def\4{{\sst{(4)}}}
\def\5{{\sst{(5)}}}
\def\6{{\sst{(6)}}}
\def\7{{\sst{(7)}}}
\def\8{{\sst{(8)}}}
\def\n{{\sst{(n)}}}
\def\oo{{\"o}}
\def\hA{\hat{\cal A}}
\def\ns{{\sst {\rm NS}}}
\def\rr{{\sst {\rm RR}}}
\def\tH{{\widetilde H}}
\def\tB{{\widetilde B}}
\def\cA{{\cal A}}
\def\cF{{\cal F}}
\def\tF{{\wtd F}}
\def\Z{\rlap{\sf Z}\mkern3mu{\sf Z}}
\def\ep{{\epsilon}}
\def\IIA{{\rm IIA}}
\def\IIB{{\rm IIB}}
\def\ads{{\rm AdS}}
\def\R{\rlap{\rm I}\mkern3mu{\rm R}}
\def\mapright#1{\smash{\mathop{-\!\!\!-\!\!\!-\!\!\!-\!\!\!-\!\!\!
             \longrightarrow}\limits^{#1}}}

\def\Ei{{\hbox{Ei}}}
\def\Ci{{\hbox{Ci}}}
\def\Si{{\hbox{Si}}}

\newcommand{\ho}[1]{$\, ^{#1}$}
\newcommand{\hoch}[1]{$\, ^{#1}$}
\newcommand{\bea}{\begin{eqnarray}} 
\newcommand{\eea}{\end{eqnarray}} 
\newcommand{\ra}{\rightarrow}
\newcommand{\lra}{\longrightarrow}
\newcommand{\Lra}{\Leftrightarrow}
\newcommand{\aap}{\alpha^\prime}
\newcommand{\bp}{\tilde \beta^\prime}
\newcommand{\tr}{{\rm tr} }
\newcommand{\Tr}{{\rm Tr} } 
\newcommand{\NP}{Nucl. Phys. }
\newcommand{\tamphys}{\it Center for Theoretical Physics,
Texas A\&M University, College Station, TX 77843}

\newcommand{\upenn}{\it Department of Physics and Astronomy,\\ University
of Pennsylvania, Philadelphia, PA 19104}

\newcommand{\brussels}{\it Physique Th\'eorique et Math\'ematique, 
Universit\'e Libre de Bruxelles,\\ Campus Plaine C.P. 231, B-1050
Bruxelles, Belgium} 

\newcommand{\auth}{J. F. V\'azquez-Poritz}

\begin{document}
\begin{flushright}
ULB-TH/01-37\\
November  2001\\
\hfill{\bf hep-th/0111229}\\
\end{flushright}

%\vspace{10pt}

\begin{center}

{\large {\bf Gravity-Trapping Domain Walls From Resolved Branes}}

\vspace{20pt}

\auth

\vspace{10pt}
% {\hoch{\dagger}\brussels}
\brussels\\
\vspace{10pt}

\vspace{30pt}

\underline{ABSTRACT}
\end{center}

All previous Randall-Sundrum type models have required a $Z_2$ identification 
source which does not have a known string theoretic origin. We show that the
near-horizon of various resolved branes on an Eguchi-Hanson instanton
dimensionally reduce to a five-dimensional domain wall that traps gravity,
without an additional delta-function source. This brings us substantially
closer to embedding infinite extra dimensions in M-theory. Also, this provides
us with a brane world model for a strongly-coupled Yang-Mills field theory
with quark-antiquark charge screening at finite separation distance.

{\vfill\leftline{}\vfill
\vskip 10pt \footnoterule {\footnotesize \hoch{1} This work is supported
in part by the Francqui Foundation (Belgium), the Actions de Recherche
Concert{\'e}es of the Direction de la Recherche Scientifique - Communaut\'e
Francaise de Belgique, IISN-Belgium (convention 4.4505.86).
 
\vskip  -12pt} \vskip   14pt
%{\footnotesize
%        \hoch{2}        Research supported in part by DOE 
%grant DOE-FG03-95ER40917 \vskip -12pt}  \vskip  14pt
}

\pagebreak
\setcounter{page}{1}

%\tableofcontents
%\addtocontents{toc}{\protect\setcounter{tocdepth}{2}}
%\newpage

\section{Introduction}

Randall and Sundrum \cite{randall1,randall2} have shown that, with fine
tuned brane tension, a flat 3-brane embedded in $AdS_5$ can have a single
massless bound state. Four-dimensional gravity is recovered at low-energy
scales. Five-dimensional domain walls which localize gravity may arise from a 
sphere reduction from ten or eleven dimensions, as the near-horizon of extremal
$p$-branes \cite{cvetic}. However, in order to localize a massless graviton
state, one must add a delta-function source which comes about from
imposing $Z_2$ symmetry. As of yet, this source has no known origin in the ten
or eleven-dimensional theory. This has prevented gravity-trapping domain walls
from fully arising in the context of string theory. 

From the type IIB perspective, in which the brane world scenario could arise 
from the near-horizon of a D3-brane, it has been proposed that the $Z_2$ 
identification source could arise from D7-branes \footnote{It has also been
proposed that the $AdS_4$ brane in $AdS_5$ can be realized in ten
dimensions as D5-branes in the near-horizon of D3-branes \cite{randall3}.}. If
the spherical symmetry of the transverse space is relaxed, one could wrap
D7-branes around a 4-cycle of the five-dimensional compact space
\cite{duff}. However, the D7-charge would need to be canceled. One candidate
for this is an orbifold charge associated with the base of an elliptically
fibered F-theory Calabi-Yau (complex) 4-fold \cite{chan}. In this case, the
dimensions transverse to the D3-brane are necessarily compact. 

In the present paper, we will also relax the spherical symmetry of the
transverse space. Our goal is to find a $p$-brane interpretation of a 
Randall-Sundrum brane world in an infinite extra dimension, without having to
add a $Z_2$ identification source. In particular, we will consider the
replacement of the standard flat transverse space by a smooth space of special
holonomy\footnote{A Ricci-flat space with  fewer covariantly constant
spinors.}, preserving a fraction of the original supersymmetry. In general,
however, the resulting solution is singular.

In certain cases, deformations of the standard brane solutions can have the 
effect of "resolving" such singularities. These deformations are the result of
an additional flux on a Ricci-flat space transverse to the brane. Since this
resolution can break additional supersymmetry, such non-singular solutions
may serve as viable gravity duals of strongly-coupled Yang-Mills field
theories with less than maximal supersymmetry
\cite{malda,kleb,strass,grana,gub,clp,cglp}. 

In the IR (small $r$) regime, the bound-state spectrum of a minimally-coupled
scalar (dilaton) corresponds to the spectrum of the operator Tr $F^2$ on the
gauge theory side, thus providing information about the glueball spectrum. The
spectrum of a minimally-coupled scalar in the background of a resolved heterotic
5-brane on an Eguchi-Hanson instanton is continuous above a mass gap. Boundary
conditions eliminate bound states of non-positive mass \cite{cglp}. This type
of spectrum is associated with a complete screening of charges in a
quark-antiquark pair at a finite separation distance.

A minimally-coupled scalar may also be associated with the graviton
fluctuations polarized along the brane worldvolume. In this case, with
appropriate boundary conditions, there is a massless bound state as well as a
continuous spectrum above a mass gap. This provides a ten-dimensional
supergravity solution whose dimensional reduction is a gravity-trapping domain
wall. It should be emphasized that no additional delta-function source from
$Z_2$ identification is needed to localize the massless graviton state. The
above also applies for a D4-brane on $S^1 \times M_4$ and a D3-brane on $T^2
\times M_4$, where $M_4$ is the Eguchi-Hanson instanton.

This paper is organized as follows. In section 2, we show how a resolved
heterotic 5-brane, D4-brane and D3-brane on an Eguchi-Hanson instanton give rise
to the same five-dimensional domain wall. In section 3, we show that this domain
wall has a bound massless graviton state with a continuous Kaluza-Klein spectrum
above a mass gap. We derive the modified Newtonian potential. In section 4, we
provide concluding remarks.

\section{Five-dimensional domain wall from resolved branes on\\ Eguchi-Hanson
instanton}

We will briefly describe the various resolved branes on an Eguchi-Hanson
instanton. Derivations and details for the case of the heterotic 5-brane and
D4-brane are provided in \cite{clp} and for the D3-brane in \cite{bert,fre}.

\subsection{Heterotic 5-brane}

A deformed heterotic 5-brane on an Eguchi-Hanson instanton is given by
%%%%%%%%%
$$
ds_{10}^2=H^{-1/4}(-dt^2+dx_j^2)+H^{3/4}ds_4^2,
$$
\begin{equation}
\textrm{e}^{-\phi}\ast F_{3}=d^6 x\wedge dH^{-1},\ \ \
\phi=\frac{1}{2}\log H,\ \ \ F_{(2)}=mL_{(2)},
\end{equation}
%%%%%%%
where $j=1,..,5$. The metric for the Eguchi-Hanson instanton is
%%%%%%%
\be
ds_4^2=W^{-1}dr^2+\frac{1}{4}r^2 W (d\psi+\cos \theta d\phi)^2+\frac{1}{4} r^2
d\Omega_2^2,\label{Eg}
\ee
%%%%%%%%
where
%%%%%
\be
W=1-\frac{a^4}{r^4},
\ee
%%%%%%%%
and $d\Omega_2^2=d\theta^2+\sin ^2 \theta d\phi^2$. The radial coordinate has
the range $a \le r \le \infty$. $F_{(2)}$ is provided by an Abelian $U(1)$ field
in the transverse space, and acts as a Chern-Simons type term:
%%%%
\be
dF_{(3)}=\frac{1}{2}F_{(2)}\wedge F_{(2)}.
\ee
%%%%
In order to satisfy the equations of motion,
%%%%%%%
\be
\square H=-\frac{1}{4}m^2 L_{(2)}^2.\label{eqH}
\ee
%%%%%%%
We consider $H$ depending only on $r$. $L_{(2)}$ must be a self-dual harmonic
2-form, which has been found to be
%%%%%
\be
L_{(2)}=r^{-3}dr \wedge (d\psi+\cos \theta d\phi)+\frac{1}{2}r^{-2}\sin \theta
d\theta \wedge d\phi.
\ee
%%%%%%
Since this harmonic function is normalizable, the previously-mentioned singular
term in $H$ can be canceled. (\ref{eqH}) becomes
%%%%%%%
\be
\frac{1}{r^3}\partial_r r^3 W\partial_r H=-\frac{4m^2}{r^8},
\ee
%%%%%%
yielding
%%%%%%%
\be
H=1+\frac{m^2+a^4b}{4a^6}{\rm log}\big(
\frac{r^2-a^2}{r^2+a^2}\big) +\frac{m^2}{2a^4 r^2}.\label{HH}
\ee
%%%%%%%
Choosing the integration constant $b=-m^2/a^4$, the logarithmic term
corresponding to a naked singularity at $r=a$ cancels, so that
%%%%%%%
\be
H=1+\frac{R^2}{r^2},\label{H}
\ee
%%%%%%%%
where $R^2=m^2/(2a^4)$. Thus, the singularity has been resolved. This solution
provides a gravity dual of $N=2$ six-dimensional field theory. In the IR (small
$r$) regime, the bound-state spectrum of a minimally-coupled scalar
(dilaton) corresponds to the spectrum of the operator Tr $F^2$ on the gauge
theory side, thus providing information about the glueball spectrum. The
spectrum of a minimally-coupled scalar in the background of a resolved heterotic
5-brane on an Eguchi-Hanson instanton is continuous above a mass gap. Boundary
conditions eliminate non-positive bound states. Such a spectrum is associated
with a complete screening of charges in a quark-antiquark pair at a finite
separation distance which is inversely proportional to the mass gap
\cite{brand2}.

We will consider the five-dimensional domain wall resulting from a reduction
over the angular coordinates of the transverse space and two coordinates on the
world volume. We can use the following Ansatz for the reduction on $S^n$
\cite{cvetic}:
%%%%%
\be
ds_D^2={\rm e}^{-2\alpha \phi}ds_d^2+g^{-2}{\rm e}^{\frac{2(d-2)}{n}\alpha \phi}
d\Omega_n^2,
\ee
%%%%%
where
%%%%%
\be
\alpha=-\sqrt{\frac{n}{2(d-2)(D-2)}}.
\ee
%%%%%
A reduction on the $\psi$, $\theta$ and $\phi$ coordinates yields a BPS domain
wall in $D=7$ gauged supergravity, which preserves 1/4 of the original
supersymmetry. Toroidally reducing over $x_4$ and $x_5$ does not change the
supersymmetry of the solution. Thus, reducing the resolved heterotic 5-brane
over $T^3 \times S^2$ (5-brane wrapped on $T^2$) yields, in the decoupling limit
($H\rightarrow R^2/r^2$), the following five-dimensional domain wall solution:
%%%%%%%
\be
ds_5^2=\big( \frac{r}{R}\big)^{4/3}W^{1/3}(-dt^2+dx_i^2)+\big(
\frac{R}{r}\big) ^{2/3}W^{-2/3}dr^2,\label{metric}
\ee
%%%%%%%
where $i=1,2,3$. If we make the coordinate transformation
%%%%%%%
\be
\frac{r}{a}={\rm cosh}^{1/2}(kz),
\ee
%%%%%%
then the metric (\ref{metric}) can be written in the conformally-flat frame as
%%%%%%%
\be
ds_5^2={\rm sinh}^{2/3}(kz)(-dt^2+dx_i^2+dz^2).\label{metric2}
\ee
%%%%%%%
$z=0$ corresponds to $r=a$. This differs from most brane world scenarios arising
from $p$-brane origins, in that the $z$ and $r$ coordinates are usually
inversely related. A result of this difference is shown in Figure 1. 
%%%%%%%%%
\begin{figure}
   \epsfxsize=4.0in
   \centerline{\epsffile{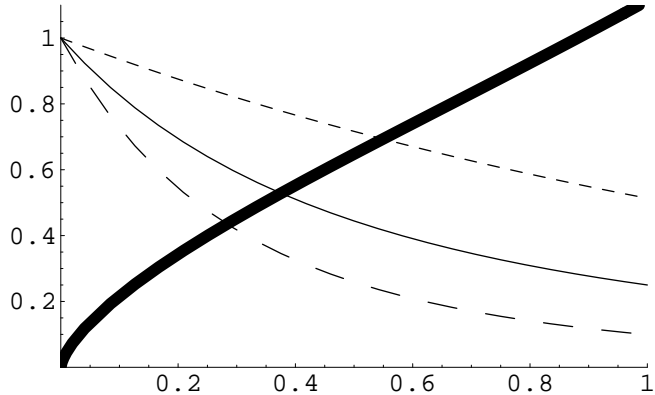}}
   \caption[FIG. \arabic{figure}.]{Conformal factors for various
five-dimensional domain walls}
\end{figure}
%%%%%%%%%
The conformal factors arising from the near-horizon of a D3-brane (normal
line), D4 or M5--brane (large dashing) and D5-brane or M5/M5-brane
intersection \cite{cvetic} (small dashing) are a positive finite value at
$z=0$ and decrease asymptotically to zero as $z$ gets larger. On the other
hand, the conformal factor for the near-horizon of the resolved heterotic
5-brane on an Eguchi-Hanson instanton (bold line) vanishes at $z=0$ and
increases as $z$ gets larger. As we will show, the result of this behavior is
an effective potential that has an infinite well at $z=0$ and is capable of
having a bound massless graviton state.

\subsection{D4-brane}

The metric of the D4-brane solution of type IIA supergravity is given by
%%%%%%%%%
\be
ds_{10}^2=H^{-3/8}(-dt^2+dx_k^2)+H^{5/8}ds_5^2,
\ee
%%%%%%%%
where $k=1,..,4$. We take the transverse space to be $M_5=M_4 \times S^1$, and
we can take $M_4$ to be the Eguchi-Hanson instanton. This transverse space can
support a normalizable self-dual harmonic 2-form, as in the case of the
heterotic 5-brane. Thus, as before, the singularity can be resolved and the
corresponding $H$ is given by (\ref{H}). Reducing the resolved D4-brane on the
angular coordinates in $M_5$ as well as the worldvolume coordinate $x_4$
($T^3\times S^2$, with the D4-brane wrapped on $S^1$), we obtain in the
decoupling limit a regular domain wall in $D=5$, with the same metric as in the
case of the heterotic 5-brane, given by (\ref{metric}). Again, we express the
metric in the conformally-flat frame given by (\ref{metric2}).

\subsection{D3-brane}

The metric of the D3-brane solution of type IIB supergravity is given by
%%%%%%%%%
\be
ds_{10}^2=H^{-1/2}(-dt^2+dx_i^2)+H^{1/2}ds_6^2,
\ee
%%%%%%%%
where $i=1,2,3$. We take the transverse space to be \cite{bert,fre}
%%%%%%
\be
ds_6^2=dzd\bar z+ds_4^2,
\ee
%%%%%%
where $ds_4^2$ is the metric of the Eguchi-Hanson instanton given by
(\ref{Eg}) and the complex coordinate $z=x_4+ix_5$. The self-dual five-form
field strength supporting the D3-brane must obey the equation of motion
%%%%%%%
\be
d\ast F_{(5)}^{RR}=-F_{(3)}^{NS}\wedge F_{(3)}^{RR},
\ee
%%%%%%%%
where a deforming three-form flux provides the Chern-Simons type term. This
three-form can be expressed in terms of a harmonic function $\gamma$ in $R^2
\sim C$ and an anti-self-dual harmonic two form in the Eguchi-Hanson instanton
space. The details of this are provided in \cite{bert,fre}. As long as there is
no
source term from a boundary action, we can consider $H$ to depend only on
$r$. In this case, $H$ has the form (\ref{HH}) with $m^2\rightarrow
a^4/(2\pi^2)$. As in the previous cases, the singularity can be resolved and the
corresponding $H$ is given by (\ref{H}).

Reducing the resolved D3-brane on the angular coordinates in the Eguchi-Hanson
instanton space as well as $x_4$ and $x_5$ ($T^3\times S^2$), we obtain in the
decoupling limit a regular domain wall in $D=5$, with the same metric as in the
two previous cases, given by (\ref{metric}). As before, we express this metric
in the conformally-flat frame given by (\ref{metric2}).

\section{Localization of graviton}

The equation of motion for a graviton fluctuation is
%%%%
\be
\partial_M \sqrt{-g}g^{MN}\partial_N \Phi=0.
\ee
%%%%%
We take $\Phi=\phi(z)M(t,x_i)$, where $\square_{(4)}M=m^2 M$ and $\square_{(4)}$
is the Laplacian on $t,x_i$. For the background (\ref{metric2}) the radial wave
equation is
%%%%%
\be
-\frac{1}{{\rm sinh}(kz)}\partial_z {\rm sinh}(kz)\partial_z
\phi=m^2\phi.\label{wave}
\ee
%%%%%%%
%%%%%%%%%
\begin{figure}
   \epsfxsize=4.0in
   \centerline{\epsffile{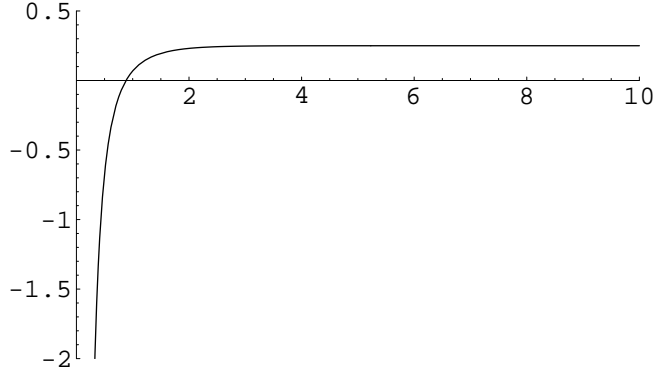}}
   \caption[FIG. \arabic{figure}.]{$V(z)$ versus $kz$}
\end{figure}
%%%%%%%%%
With the wave function transformation
%%%%%%
\be
\phi={\rm sinh}^{-1/2}(kz)\psi,
\ee
%%%%%%
the wave equation (\ref{wave}) becomes
%%%%%%%
\be
-\partial_z^2 \psi+V(z)\psi=m^2\psi,
\ee
%%%%%%%
where the effective potential 
%%%%%
\be
V(z)=\frac{k^2}{4}(1-{\rm sinh}^{-2}(kz)).
\ee
%%%%
As shown in Figure 2, even without introducing an additional delta-function
source, $V(z)$ is similar to the effective potential for the case
of a gravity-trapping domain wall derived from a D5-brane on a
standard flat transverse space \cite{cvetic}, which is 
%%%%%%
\be
\tilde{V}(r)=\frac{k^2}{4}-k\delta (z).
\ee
%%%%%%%
The graviton spectrum has a mass gap of
%%%%%%%%
\be
m_{{\rm gap}}^2=V(z\rightarrow \infty)=\frac{k^2}{4}.
\ee
%%%%%%%%
This separation between the massless graviton and the massive modes better
ensures that we have a well-defined effective field theory \cite{brand}.
Other gravity-trapping domain walls whose spectrum has a mass gap include that
associated with a D5-brane reduced on $T^2 \times S^3$ \cite{cvetic} and a
D3-brane distributed over a disk \cite{brand}. In all cases, the mass gap
results, in the $p$-brane perspective, from the harmonic function $H$
supported by a four-dimensional transverse space. 
%%%%%%%%%
\begin{figure}
   \epsfxsize=4.0in
   \centerline{\epsffile{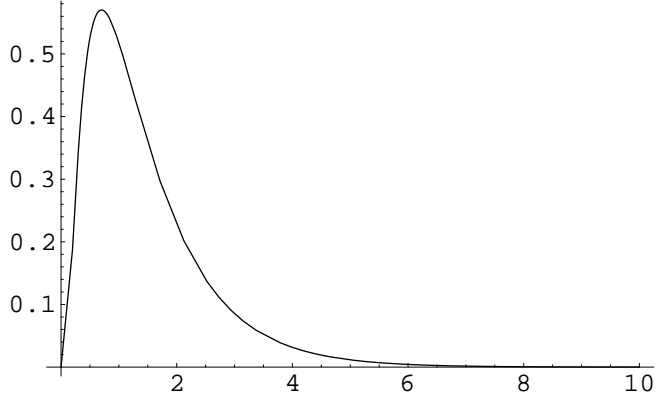}}
   \caption[FIG. \arabic{figure}.]{$\psi_0(z)$ versus $kz$}
\end{figure}
%%%%%%%%%

The wave function solution is 
%%%%%%
$$
\psi_m=N_m {\rm sinh}^{-i\gamma}(kz)F[\frac{1}{4}-i\frac{\sqrt{3}}{4}+
i\frac{\gamma}{2},\frac{1}{4}+i\frac{\sqrt{3}}{4}+i\frac{\gamma}{2},
1+i\gamma,-{\rm sinh}^{-2}(kz)]+
$$
\be
M_m {\rm sinh}^{i\gamma}(kz)F[\frac{1}{4}-i\frac{\sqrt{3}}{4}-
i\frac{\gamma}{2},\frac{1}{4}+i\frac{\sqrt{3}}{4}-i\frac{\gamma}{2},
1-i\gamma,-{\rm sinh}^{-2}(kz)],\label{sol}
\ee
%%%%%%%
where $F[a,b,c,y]$ is the solution to the hypergeometric equation
%%%%%%%
\be
y(1-y)F''(y)+[c-(a+b+1)y]F'(y)-abF(y)=0,
\ee
%%%%%%%%
and $\gamma =\sqrt{(m/k)^2-1/4}$. In order for the solution not to blow up at
large $z$, we set $M_m=0$. For the massless graviton mode, the remaining
solution given in (\ref{sol}) reduces to
%%%%%%
\be
\psi_0=N_0 {\rm sinh}^{1/2}(kz)\cos [\frac{\sqrt{3}}{2}{\rm arcsinh} ({\rm
sinh}^{-1}(kz))].
\ee
%%%%%%
The trapping of gravity requires that the wave function is normalizable
\cite{cvetic}, in order to obtain a finite leading-order contribution to the
gravitational field on the brane world. We numerically find that $\int dz
|\psi_0|^2$ is finite, and the normalization factor $N_0=1.07077\sqrt{k}$. As
shown in Figure 3, the peak of $\psi_0(z)$ is located at $kz\approx .7 \equiv
kz_0$. We propose that this is the location of a four-dimensional brane
world \footnote{In previous Randall-Sundrum type models, one needed an
additional delta-function source in order to localize gravity to the
brane. This source was the result of patching together five-dimensional
supergravity solutions in a discontinuous manner by taking $z\rightarrow
c+|z|$, where $c$ is a positive constant. If we extrapolate our solution to
$z<0$, there is already $Z_2$ symmetry about $z=0$, which implies that there
is a brane world at $z=-z_0$.}.

The massive modes oscillate in the bulk, as shown in Figure 4.
%%%%%%%%%
\begin{figure}
   \epsfxsize=4.0in
   \centerline{\epsffile{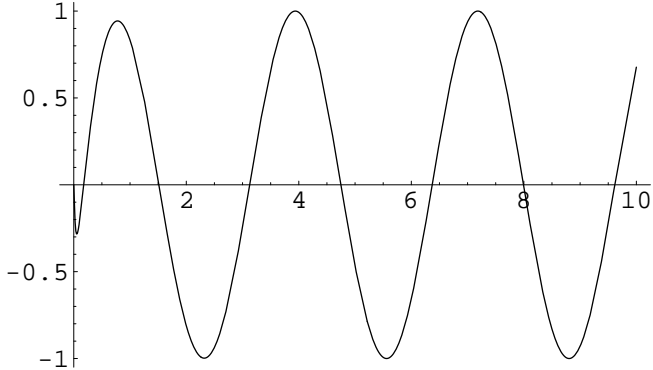}}
   \caption[FIG. \arabic{figure}.]{Re $\psi_m(z)$ versus $kz$ for $m=2k$}
\end{figure}
%%%%%%%%%
For large $kz$ the massive wave functions given by (\ref{sol}) are approximately
%%%%%
\be
\psi_m \sim N_m {\rm e}^{-i\gamma kz}.
\ee
%%%%%%
This can be normalized via the delta-function orthogonality condition to yield
$N_m=1/\sqrt{\pi \gamma k}$.

The Newtonian gravitational potential between masses $M_1$ and $M_2$ can be
estimated\\ by \cite{brand}
%%%%%%%
\be
U(r)\sim \frac{G_4 M_1 M_2}{r}+\frac{G_5 M_1
M_2}{r}\int_{k^2/4}^{\infty}d(m^2) |\psi_m(z_0)|^2 {\rm e}^{-mr}.\label{pot}
\ee
%%%%%%%%
The four and five-dimensional Newton constants are related by $G_4=kG_5$. To
find the leading-order correction, we consider $y\equiv m-k/2 << k$. In this
limit, $|\psi_m(z_0)|^2 \approx .543N_m^2$. Inserting this into (\ref{pot}), we
find that
%%%%%%
\be
U(r)\sim \frac{G_4 M_1 M_2}{r}\Big( 1+\frac{.543}{\sqrt{\pi}}\frac{{\rm
e}^{-\frac{1}{2}kr}}{\sqrt{kr}}\Big).
\ee
%%%%%
The Yukawa-like factor ${\rm e}^{-\frac{1}{2}kr}$ in the sub-leading term
reflects the presence of a mass gap $\frac{1}{2}k$ separating the zero-energy
bound state and the massive Kaluza-Klein continuum.

\section{Conclusions}

We have shown that the near-horizon of various resolved branes on an
Eguchi-Hanson instanton dimensionally reduces to a five-dimensional domain
wall that traps gravity. Thus, this Randall-Sundrum type brane
world arises from a $p$-brane origin without an additional delta-function
source from imposing $Z_2$ identification. This brings us substantially closer
to embedding infinite extra dimensions in M-theory. Also, this provides us
with a brane world model for a strongly-coupled Yang-Mills field theory with
quark-antiquark charge screening at finite separation distance.

The Eguchi-Hanson manifold is the simplest example in a class of spaces that are 
asymptotically locally Euclidean (ALE). It may be interesting to see if resolved
branes on generic smooth ALE manifolds dimensionally reduce to gravity-trapping
domain walls, or whether this is a property of the Eguchi-Hanson instanton in
particular.

Also, the question arises as to whether the zero-modes of other bulk fields
can be localized to this brane world in a similar manner as for the graviton.

Lastly, as mentioned in the Introduction, a minimally-coupled scalar in the bulk
can correspond to either the operator Tr $F^2$ on the gauge theory side
(AdS/CFT perspective) or to the graviton fluctuations polarized along the brane
worldvolume (Randall-Sundrum perspective). In the IR (small $r$) regime the
spectrum of the glueballs and worldvolume gravitons are expected to be
identical \cite{brand2}. For a confining gauge theory, there is a discrete
spectrum of bound states \cite{cglp}. This would not be desirable from the
Randall-Sundrum perspective unless the zero-mode state corresponding to the
four-dimensional graviton dominates over the massive bound states, within
experimental bounds.

\section*{Acknowledgments}

We are grateful to Marc Henneaux and Malcolm Fairbairn for useful discussions.


\begin{thebibliography}{99}

\bm{randall1} L. Randall and R. Sundrum, {\sl An alternative to
compactification}, Phys. Rev. Lett. 83 (1999) 4690-4693, hep-th/9906064.

\bm{randall2} L. Randall and R. Sundrum, {\sl A large mass hierarchy from
a small extra dimension}, Phys. Rev. Lett. 83 (1999) 3370, hep-ph/9905221.

\bm{cvetic} M. Cveti\v{c}, H. L\"{u} and C.N. Pope, {\sl Domain walls
with localised gravity and domain-wall/QFT correspondence}, Phys. Rev. D63
(2001) 086004, hep-th/0007209.

\bm{duff} M. Cveti\v{c}, M.J. Duff, J.T. Liu, H. L\"{u}, C.N. Pope and
K.S. Stelle, {\sl Randall-Sundrum Brane Tensions}, Nucl. Phys. B605
(2001) 141-158, hep-th/0011167.

\bm{randall3} A. Karch and L. Randall, {Localized gravity in string theory},
Phys. Rev. Lett. 87 (2001) 061601, hep-th/0105108.

\bm{chan} C.S. Chan, P.L. Paul and H. Verlinde, {\sl A note on warped string
compactification}, Nucl. Phys. B581 (2000) 156, hep-th/0003236.

\bibitem{malda} J. Maldacena, {\sl The large $N$ limit of
superconformal field theories and supergravity},
Adv. Theor. Math. Phys. {\bf 2} (1998) 231, hep-th/9711200.

\bibitem{kleb} I.R. Klebanov and A.A. Tseytlin, {\sl Gravity duals of
supersymmetric SU(N) $\times$ SU(N+m) gauge theories}, Nucl. Phys. B578
(2000) 123, hep-th/0002159.

\bibitem{strass} I.R. Klebanov and M.J. Strassler, {\sl Supergravity and a
confining gauge theory: duality cascades and $\Xi$SB-resolution of naked
singularities}, JHEP 0008:052, 2000, hep-th/0007191.

\bibitem{grana} M. Gra\~{n}a and J. Polchinski, {\sl Supersymmetric
three-form flux perturbations on $AdS_5$}, Phys. Rev. D63 (2001) 026001,
hep-th/0009211. 

\bibitem{gub} S. Gubser, {\sl Supersymmetry and F-theory realization of the
deformed conifold with three-form flux}, hep-th/0010010.

\bibitem{clp} M. Cveti\v{c}, H. L\"{u} and C.N. Pope, {\sl Brane
Resolution Through Transgression}, Nucl. Phys. B600 (2001) 103-132,
hep-th/0011023.

\bibitem{cglp} M. Cveti\v{c}, G.W. Gibbons, H. L\"{u} and C.N. Pope, {\sl
Ricci-flat metrics, harmonic forms and brane resolutions}, hep-th/0012011.

\bm{bert} M. Bertolini, V.L. Campos, G. Ferretti, P. Fr\'e, P. Salomonson and
M. Trigiante, {\sl Supersymmetric 3-branes on smooth ALE manifolds with flux},
hep-th/0106186.

\bm{fre} P. Fr\'e, {BPS D3-branes on smooth ALE manifolds}, hep-th/0110281.

\bm{brand} A. Brandhuber and K. Sfetsos, {\sl Non-standard compactification with
mass gaps and Newton's law}, JHEP 9910 (1999) 013, hep-th/9908116.

\bm{brand2} A. Brandhuber and K. Sfetsos, {\sl An $N=2$ gauge theory and its
supergravity dual}, Phys. Lett. B488 (2000) 373-382, hep-th/0004148.


\end{thebibliography}
\end{document}